\documentclass[12pt]{article}

\usepackage{amssymb}
\usepackage{amsmath}
\usepackage{latexsym}
\usepackage{yfonts}

\catcode`@=11
\renewcommand{\section}{\@startsection{section}{1}{0pt}{\medskipamount}
{\medskipamount}{\large\bf}}
\numberwithin{equation}{section}
\catcode`@=12

\def\beq{\begin{eqnarray}}    
\def\eeq{\end{eqnarray}}      

\def\pa{\partial}                       

\def\={\ =\ }

\begin{document}

\begin{center}

{\Large\bf Quantum Gravity  and  background field formalism}

\vspace{18mm}

{\large 
Peter M. Lavrov$^{(a, b)} \footnote{E-mail: lavrov@tspu.edu.ru}$,\;
}

\vspace{8mm}

\noindent  ${{}^{(a)}} ${\em
Tomsk State Pedagogical University,\\
Kievskaya St.\ 60, 634061 Tomsk, Russia}

\noindent  ${{}^{(b)}} ${\em
National Research Tomsk State  University,\\
Lenin Av.\ 36, 634050 Tomsk, Russia}

\vspace{20mm}

\begin{abstract}
\noindent
We analyze the problem of general covariance for quantum gravity
theories in the background field formalism with respect to gauge fixing procedure. We prove that the
background effective action is not invariant under
general coordinate transformations of background metric tensor 
in non-linear gauges.

\end{abstract}

\end{center}

\vfill

\\

\newpage

\section{Introduction}
\noindent The background field formalism \cite{DeW,AFS,Abbott} is one
of the most popular method for quantum studies  and calculations in
gauge theories because it allows to work with  the effective action
invariant under the gauge transformations of background fields and
to reproduce all usual physical results by choosing a special
background gauge fixing condition. Various aspects of quantum
properties of Yang-Mills theories have been successfully studied in
this technique
\cite{'tH,K-SZ,GvanNW,CMacL,Gr,Barv,FT,BLT-YM,Lav}.

A classical action of all quantum gravity theories obeys the property of invariance under
general coordinate transformations and can be considered as an example of gauge theory with
closed gauge algebra and with structure coefficients independent on fields (the metric tensor).
For such kind of theories the quantization can be performed in
the form of Faddeev-Popov procedure \cite{FP}. Because similarity between
Yang-Mills theories and gravity theories as gauge theories it seems naturally to apply the
background field formalism being very successfully in the case of Yang-Mills fields to study their quantum properties. We are going  to consider more detailed the problem of general covariance
of the background effective action for quantum gravity theories with respect to gauge fixing procedure.

In the present paper we analyze the general covariance of the effective action for any initial
classical gravity action in  the background field formalism. Application of this method
to Yang-Mills theories gives rise two important advantages of the effective action:
gauge invariance and gauge independence on its extremals. In the case of Quantum Gravity
formulated within the background field formalism we confirm the property of gauge independence
of the effective action on its extremals for all admissible gauges but we point out that the gauge invariance
of the background effective action is supported by linear vector gauges only.

The paper is organized as follows. In Section 2 we fix notations and represent
arbitrary gravity theories in  the background field formalism in any admissible gauges
to confirm gauge independence of vacuum functional on gauge conditions and as a consequence
the same property of effective action on its extremals. In Section 3 the general covariance
of vacuum functional is analyzed. It is found that the general covariance
can be arrived at the propositions  concerning validity of  tensor transformations
of gauge fixing functions and its linear dependence on quantum gravitational fields. 
In Section 4
the simplest case of gauge fixing condition is considered to check previous assumptions. It is shown that the standard choice of gauge fixing functions satisfies the required propositions.
In Section 5 concluding discussions are given.

In the paper the DeWitt's condensed notations are used \cite{DeWitt}.
We employ the notation $\varepsilon(A)$ for the
Grassmann parity and the ${\rm gh}(A)$ for the ghost number of any quantity $A$ .
All functional derivatives  are taken from the left.  The functional right
derivatives with respect to fields are marked by special
symbol $"\leftarrow"$.

\section{Background field formalism  for Quantum Gravity: gauge independence}

\noindent Our starting point is an arbitrary action of a Riemann's
metric, $S_0=S_0(g)$, $g=\{g_{\mu\nu}\}$ invariant under the general
coordinate transformations, \footnote{Standard examples are Einstein
gravity, $S_0(g)=\kappa^{-2}\int dx \sqrt{-g}\;\!R$, and $R^2$
gravity, $S_0(g)=\int dx \sqrt{-g}\;(\lambda_1 R^2+
\lambda_2R^{\mu\nu}R_{\mu\nu}+\kappa^{-2}R)$.}
\beq
\label{A1}
{x'}^{\mu}=f^{\mu}(x)\;\rightarrow \;x^{\mu}=x^{\mu}(x'),\quad
g_{\mu\nu}\;\rightarrow\; g'_{\mu\nu}(x')=
g_{\alpha\beta}(x)\frac{\pa x^{\alpha}}{\pa {x'}^{\mu}}\frac{\pa
x^{\beta}}{\pa {x'}^{\nu}}.
\eeq
In the infinitesimal form the
transformations (\ref{A1}) read
\beq \label{A2}
{x'}^{\mu}=x^{\mu}+\omega^{\mu}(x)\;\rightarrow
\;x^{\mu}={x'}^{\mu}-\omega^{\mu}(x'),\quad g_{\mu\nu}\;\rightarrow\;
g'_{\mu\nu}(x)=g_{\mu\nu}(x)+\delta_{\omega} g_{\mu\nu}(x),
\eeq
where
\beq
\label{A3}
\delta_{\omega}g_{\mu\nu}(x)=-\omega^{\sigma}(x)\pa_{\sigma}g_{\mu\nu}(x)-
g_{\mu\sigma}(x)\pa_{\nu}\omega^{\sigma}(x)-g_{\sigma\nu}(x)\pa_{\mu}\omega^{\sigma}(x).
\eeq
The action $S_0(g)$ is invariant under the transformations
(\ref{A3})
\beq
\label{A4}
\int dx \frac{\delta S_0(g)}{\delta
g_{\mu\nu}(x)}\delta_{\omega} g_{\mu\nu}(x)=0.
\eeq
For any tensor fields $A_{\mu}(x), A^{\mu}(x), A_{\mu\nu\lambda}(x)$ of types
$(0,1), (1,0), (0,3), (1,2)$,  respectively, on a given manifold
the infinitesimal form of general coordinate transformations is
\beq
\label{A5}
&&\delta_{\omega}A_{\mu}(x)=-\omega^{\sigma}(x)\pa_{\sigma}A_{\mu}(x)-
A_{\sigma}(x)\pa_{\mu}\omega^{\sigma}(x),\\
\label{A6}
&&\delta_{\omega}A^{\mu}(x)=-\omega^{\sigma}(x)\pa_{\sigma}A^{\mu}(x)+
A^{\sigma}(x)\pa_{\sigma}\omega^{\mu}(x),\\
\nonumber
&&\delta_{\omega}A_{\mu\nu\lambda}(x)=-\omega^{\sigma}(x)\pa_{\sigma}A_{\mu\nu\lambda}(x)
-A_{\mu\nu\sigma}(x)\pa_{\lambda}\omega^{\sigma}(x)-\\
\label{A7}
&&\qquad\qquad\quad\;\;-A_{\mu\sigma\lambda}(x)\pa_{\nu}\omega^{\sigma}(x)
-A_{\sigma\nu\lambda}(x)\pa_{\mu}\omega^{\sigma}(x),
\\
\nonumber
&&\delta_{\omega}A_{\mu\nu}^{\lambda}(x)=-\omega^{\sigma}(x)\pa_{\sigma}A_{\mu\nu}^{\lambda}(x)+A^{\sigma}_{\mu\nu}(x)\pa_{\sigma}\omega^{\lambda}(x)
-\\
\label{A7a}
&&\qquad\qquad\quad\;\;-A_{\mu\sigma}^{\lambda}(x)\pa_{\nu}\omega^{\sigma}(x)
-A_{\sigma\nu}^{\lambda}(x)\pa_{\mu}\omega^{\sigma}(x).
\eeq.

Let us represent the
transformations (\ref{A3}) in the form
\beq
\label{A8}
\delta_{\omega}
g_{\mu\nu}(x)=\int dy R_{\mu\nu\sigma}(x,y;g)\omega^{\sigma}(y),
\eeq
where
\beq
\label{A9}
R_{\mu\nu\sigma}(x,y;g)=-\delta(x-y)\pa_{\sigma}g_{\mu\nu}(x)-
g_{\mu\sigma}(x)
\pa_{\nu}\delta(x-y)-g_{\sigma\nu}(x)\pa_{\mu}\delta(x-y)
\eeq
can
be considered as the generators of gauge transformations of the
metric tensor $g_{\mu\nu}$ with gauge parameters $\omega^{\sigma}(x)$.
The algebra of gauge transformations has the following form
\beq
\nonumber &&\int du \Big(\frac{\delta
R_{\mu\nu\sigma}(x,y;g)}{\delta
g_{\alpha\beta}(u)}R_{\alpha\beta\gamma} (u,z;g)-\frac{\delta
R_{\mu\nu\gamma}(x,z;g)}{\delta
g_{\alpha\beta}(u)}R_{\alpha\beta\sigma} (u,y;g)\Big)=\\
\label{A10}
&&\qquad\qquad\qquad =- \int du
R_{\mu\nu\lambda}(x,u;g)F^{\lambda}_{\sigma\gamma}(u,y,z),
\eeq
where
\beq
\label{A11}
F^{\lambda}_{\alpha\beta}(x,y,z)= \delta
(x-y)\;\delta^{\lambda}_{\beta}\pa_{\alpha}\;\delta(x-z)-
\delta(x-z)\;\delta^{\lambda}_{\alpha}\;\pa_{\beta}\;\delta (x-y)
\eeq
are structure functions of the gauge
algebra which do not depend on the metric tensor $g_{\mu\nu}$.
Therefore, any theory of gravity looks like a gauge theory with
closed gauge algebra and structure functions independent on fields
(metric tensor), i.e. as an Yang-Mills type theory.
In what follows we will omit the space - time argument $x$ of fields and
gauge parameters when this does not lead to misunderstandings
in the formulas and relations employing the DeWitt's condensed notations \cite{DeW}.
Then the relations  (\ref{A8}), (\ref{A10})  are presented in the form
\beq
\label{A12}
&&
\qquad\qquad\qquad \delta_{\omega}g_{\mu\nu}=R_{\mu\nu\sigma}(g)\omega^{\sigma},\\
\label{A13}
&&\frac{\delta
R_{\mu\nu\sigma}(g)}{\delta
g_{\alpha\beta}}R_{\alpha\beta\gamma} (g)-\frac{\delta
R_{\mu\nu\gamma}(g)}{\delta
g_{\alpha\beta}}R_{\alpha\beta\sigma}(g)=-
R_{\mu\nu\lambda}(g)F^{\lambda}_{\sigma\gamma}.
\eeq

In the background field formalism \cite{DeW,AFS}
the metric tensor  $g_{\mu\nu}$ appearing in classical action
$S_0(g)$, is replaced by  ${\bar g}_{\mu\nu}+h_{\mu\nu}$,
\beq
\label{A14}
S_0(g)\;\rightarrow \;S_0({\bar g}+h),
\eeq
where ${\bar g}_{\mu\nu}$ is considered as a background metric tensor while
$h_{\mu\nu}$ present the quantum fields as integration variables in functional integrals for
generating functionals of Green functions.

The action $\mathcal{S}_0({\bar g}+h)$ obeys obviously the gauge invariance,
\beq
\delta_{\omega} {S}_{0}({\bar g}+h)=0,\quad
\delta_{\omega} h_{\mu\nu }=R_{\mu\nu\sigma}(h)\omega^{\sigma},\quad
\delta_{\omega}{\bar g}_{\mu\nu}=R_{\mu\nu\sigma}({\bar g})\omega^{\sigma}.
\label{A15}
\eeq
The corresponding Faddeev-Popov action
$S_{FP}=S_{FP}(\phi,{\bar g})$ is written as \cite{FP}
\beq
\label{A16}
S_{FP}=S_{0}({\bar g}+h)+S_{gh}(\phi, {\bar g})+S_{gf}(\phi,{\bar g}),
\eeq
where $S_{gh}(\phi,{\bar g})$ is the ghost action
\beq
\label{A17}
S_{gh}(\phi,{\bar g})=
\int dx \sqrt{-{\bar g}}\;
{\bar C}^{\alpha}G_{\alpha}^{\beta\gamma}({\bar g},h)
R_{\beta\gamma\sigma}({\bar g}+h)C^{\sigma},
\eeq
with  the notation
\beq
\label{A18}
G_{\alpha}^{\beta\gamma}({\bar g},h)=
\frac{\delta\chi_{\alpha}({\bar g},h)}{\delta h_{\beta\gamma}}.
\eeq
The  $S_{gf}({\bar g},h)$ is the gauge fixing action
\beq
\label{A19}
S_{gf}(\phi,{\bar g})=\int dx \sqrt{-{\bar g}}\;
B^{\alpha}\chi_{\alpha}({\bar g},h).
\eeq
Here $\chi_{\alpha}({\bar g},h)$ are functions lifting the degeneracy
of the  action $S_0$,
$\phi=\{\phi^i\}$  is the set of all fields
$\phi^i=(h_{\mu\nu},B^{\alpha}, C^{\alpha}, {\bar C}^{\alpha})$
($\varepsilon(\phi^i)=\varepsilon_i$)
with the Faddeev-Popov ghost and anti-ghost fields
$ C^{\alpha}, {\bar C}^{\alpha}$ ($\varepsilon(C^{\alpha})=
\varepsilon( {\bar C}^{\alpha})=1,\;
{\rm gh}(C^{\alpha})=-{\rm gh}({\bar C}^{\alpha})=1$), respectively, and
the Nakanishi-Lautrup auxiliary fields $B^{\alpha}$
($\varepsilon(B^{\alpha})=0,\; {\rm gh}(B^{\alpha})=0$).

For any admissible choice of gauge fixing functions $\chi_{\alpha}({\bar g},h)$ the action
(\ref{A11}) is invariant under global supersymmetry (BRST symmetry) \cite{BRS1,T}, \!\!\!
 \footnote{The gravitational BRST transformations were introduced in \cite{DR-M,Stelle,TvN}}
\beq
\label{A20}
\delta_B h_{\mu\nu}=R_{\mu\nu\alpha}({\bar g}+h)C^{\alpha}\Lambda,\quad
\delta_B B^{\alpha}=0,\quad
\delta_B C^{\alpha}=-C^{\sigma}\pa_{\sigma} C^{\alpha}\Lambda,\quad
\delta_B {\bar C}^{\alpha}=B^{\alpha}\Lambda,
\eeq
where $\Lambda$ is a constant Grassmann parameter. Let us present the BRST transformations
(\ref{A20}) in the form
\beq
\label{A21}
\delta_{B}\phi^i=R^i(\phi, {\bar g})\Lambda,\quad
\varepsilon(R^i(\phi, {\bar g}))=\varepsilon_i+1,
\eeq
where
\beq
\label{A22}
R^i(\phi, {\bar g})=\big(R_{\mu\nu\sigma}({\bar g}+h)C^{\sigma},\; 0\;,
 -C^{\sigma}\pa_{\sigma} C^{\alpha}, B^{\alpha}\big).
\eeq
Introducing the gauge fixing functional $\Psi=\Psi(\phi,{\bar g})$,
\beq
\label{A23}
\Psi=\int dx \sqrt{-{\bar g}}\;{\bar C}^{\alpha}\chi_{\alpha}({\bar g},h),
\eeq
the action (\ref{A15}) rewrites as
\beq
\label{A24}
S_{FP}(\phi,{\bar g})=S_0({\bar g}+h)+\Psi(\phi,{\bar g})
{\hat R}(\phi,{\bar g}),
\qquad S_0({\bar g}+h){\hat R}(\phi,{\bar g})=0,
\eeq
where
\beq
\label{A25}
{\hat R}(\phi,{\bar g})=\int dx\;
\frac{\overleftarrow{\delta}}{\delta\phi^i}R^i(\phi, {\bar g})
\eeq
is the generator of BRST transformations.
Due to the nilpotency property of ${\hat R}$, ${\hat R}^2=0$, the BRST symmetry of $S_{FP}$
follows from the presentation (\ref{A24}) immediately,
\beq
\label{A26}
S_{FP}(\phi,{\bar g}){\hat R}(\phi,{\bar g})=0.
\eeq

The generating functional of Green functions in the background field
method is defined in the form of functional integral
\beq
\label{A27}
Z(J, {\bar g})=\int
d\phi\;\exp\Big\{\frac{i}{\hbar}\big[S_{FP}(\phi, {\bar g})+
J\phi\big]\Big\}=\exp\Big\{\frac{i}{\hbar}W(J, {\bar g})\Big\},
\eeq
where $W(J, {\bar g})$ is the generating functional of
connected Green functions. In (\ref{A27}) the notations
\beq
\label{A28}
J\phi=\int dx \sqrt{-{\bar g}}J_i(x)\phi^i(x), \quad
J_i(x)=(J^{\mu\nu}(x), J^{(B)}_{\alpha}(x), {\bar J}_{\alpha}(x), J_{\alpha})(x)
\eeq
are used and
$J_i(x)$ \big($\varepsilon(J_i(x))=\varepsilon_i,\;
{\rm gh}(J_i(x))={\rm gh}(\phi^i(x))$\big)
are external sources to fields $\phi^i(x)$.

Let $Z_{\Psi}({\bar g})$ be the vacuum functional which corresponds
to the choice of gauge fixing functional (\ref{A23}) in the presence
of external fields ${\bar g}$,
\beq
\label{A29}
&&Z_{\Psi}({\bar g})=\int
d\phi\;\exp\Big\{\frac{i}{\hbar}\big[S_0({\bar g}+h)+
\Psi(\phi,{\bar g}) {\hat R}(\phi,{\bar g})\big]\Big\}=\\
\nonumber
&&\qquad=\int
d\phi\;\exp\Big\{\frac{i}{\hbar}S_{FP}(\phi, {\bar g})\Big\}
=\exp\Big\{\frac{i}{\hbar}W_{\Psi}({\bar g})\Big\}.
\eeq
In turn, let $Z_{\Psi+\delta\Psi}$ be the vacuum functional
corresponding to a gauge fixing functional $\Psi(\phi,{\bar g})+\delta\Psi(\phi,{\bar g})$,
\beq
\label{A30}
Z_{\Psi+\delta\Psi}({\bar g})=\int
d\phi\;\exp\Big\{\frac{i}{\hbar}\big[S_{FP}(\phi, {\bar g})+
\delta\Psi(\phi,{\bar g}){\hat R}(\phi,{\bar g})\big]\Big\}.
\eeq
Here, $\delta\Psi(\phi,{\bar g})$
is an arbitrary infinitesimal odd functional
which  may  in general has  a form  differing on (\ref{A23}).
Making use of the change of variables $\phi^i$ in the form of BRST
transformations (\ref{A20}) but with replacement of the constant parameter
$\Lambda$ by the following functional
\beq
\label{A31}
\Lambda=\Lambda(\phi,{\bar g})=\frac{i}{\hbar}\delta\Psi(\phi,{\bar g}),
\eeq
and taking into account that the Jacobian of transformations is
equal to
\beq
\label{A32}
J=\exp\{-\Lambda(\phi,{\bar g}){\hat
R}(\phi,{\bar g})\},
\eeq
we find the gauge independence of the
vacuum functional\footnote{Using the finite BRST transformations one can connect the
description of any gauge theory in two arbitrary admissible gauges
\cite{LL,BLT-fin-1,BBLT-fin}.}
\beq
\label{A33}
Z_{\Psi}({\bar g})=Z_{\Psi+\delta\Psi}({\bar g}),
\eeq
so that
\beq
\label{A34}
\delta_{\Psi}Z({\bar g})=0
\;\rightarrow\; \delta_{\Psi}W({\bar g})=0.
\eeq
The property (\ref{A33}) was
a reason to omit the label $\Psi$ in the definition of generating
functionals (\ref{A27}). In deriving (\ref{A28}) the relation
\beq
\label{A35}
 (-1)^{\varepsilon_i}\frac{\pa}{\pa\phi^i}R^i(\phi, {\bar g})=0,
\eeq
was used. The property (\ref{A28})
means that due to the equivalence theorem \cite{KT}
the physical $S$-matrix does not depend on the gauge fixing.
In terms of the effective action $\Gamma(\Phi,{\bar g})$ which   is defined with the help
of Legendre transformation
\beq
\label{A36}
\Gamma(\Phi,{\bar g})=W(J,{\bar g})-J\Phi,\quad
\frac{\delta W(J,{\bar g})}{\delta J_i}=\sqrt{-{\bar g}}\;\!\Phi^i, \quad
J\Phi=\int dx \sqrt{-{\bar g}}\;\!J_i(x)\Phi^i(x),
\eeq
the property (\ref{A34}) reads
\beq
\label{A37}
\delta_{\Psi}\Gamma(\Phi,{\bar g})
\Big|_{\frac{\delta\Gamma(\Phi,{\bar g})}{\delta\Phi}=\;\!0}=
0,
\eeq
i.e. the effective action evaluated on its extremal does not depend on gauge.

\section{Gauge invariance}

\noindent
 The gauge independence of the vacuum functional for
Quantum Gravity in the background field formalism repeats the
corresponding property for the vacuum functional for Yang-Mills
theories. Moreover that vacuum
functional for Yang-Mills theories obeys additional important
invariance property under the gauge transformations of external
vector fields \cite{Abbott}. All quantum gravity theories look like
as special type of gauge theories (similar to Yang-Mills theories)
with closed gauge algebra and with structure coefficients
independent on fields. Therefore it is natural to expect invariance
of vacuum functional for Quantum Gravity in the background field
formalism under general coordinate transformations on manifolds with
an external metric tensor ${\bar g}_{\mu\nu}$ \cite{Barv}.

Consider a variation of $Z({\bar g})$ under general coordinates transformations of
external metric tensor ${\bar g}_{\mu\nu}$,
\beq
\label{B1}
\delta^{(c)}_{\omega}{\bar g}_{\mu\nu}=R_{\mu\nu\sigma}({\bar g})\;\!\omega^{\sigma}.
\eeq
Then we have
\beq
\label{B2}
\delta^{(c)}_{\omega}Z({\bar g})=
\frac{i}{\hbar}\int d\phi\big[\delta^{(c)}_{\omega}S_0({\bar g}+h)+
\delta^{(c)}_{\omega}S_{gh}(\phi,{\bar g})+
\delta^{(c)}_{\omega}S_{gf}(\phi,{\bar g})\big]
\exp\Big\{\frac{i}{\hbar}S_{FP}(\phi, {\bar g})\Big\}.
\eeq
Now, using a change of variables in the functional integral (\ref{B2}) one should try to
arrive at the relation $\delta^{(c)}_{\omega}Z({\bar g})=0$ to prove invariance
of $Z({\bar g})$ under the transformations (\ref{B1}). In the sector of fields
$h_{\mu\nu}$  the form of this transformations is dictated by the invariance property
of $S_0({\bar g}+h)$ and reads
\beq
\label{B3}
\delta^{(q)}_{\omega}h_{\mu\nu}=R_{\mu\nu\sigma}(h)\;\!\omega^{\sigma}=-
\omega^{\sigma}\pa_{\sigma}h_{\mu\nu}-
h_{\mu\sigma}\pa_{\nu}\omega^{\sigma}-h_{\sigma\nu}\pa_{\mu}\omega^{\sigma},
\eeq
so that
\beq
\label{B4}
\delta_{\omega}S_0({\bar g}+h)=0,\quad
\delta_{\omega}=(\delta^{(c)}_{\omega}+\delta^{(q)}_{\omega}).
\eeq
Notice that on this stage there exists a difference between the Yang-Mills
theories formulated in the background field formalism and Quantum Gravity theories under
consideration. The change (\ref{B3}) is just a gauge transformation of quantum fields $h_{\mu\nu}$ while in the case of Yang-Mills theories the corresponding change has the form of tensor transformations of quantum vector fields \cite{Abbott}. It is the reason for us to consider the invariance property of the action (\ref{A17}) in detail.

Next step is related with analysis of the gauge fixing action $S_{gf}(\phi,{\bar g})$ because
it depends only on three variables $h_{\mu\nu}, B^{\alpha}, {\bar g}_{\mu\nu}$ and
for two of them, $h_{\mu\nu}, {\bar g}_{\mu\nu}$, the transformation law is already defined
(\ref{B1}), (\ref{B3}). Let $\delta_{\omega}B^{\alpha}$ be at the moment
unknown transformation of fields $B^{\alpha}$. The explicit form
of $\delta_{\omega}B^{\alpha}$ should be chosen
in a such of way  to compensate the variation of $S_{gf}(\phi,{\bar g})$
caused by transformations ${\bar g}_{\mu\nu}$ and $h_{\mu\nu}$. In the case of Yang-Mills
theories it can be done with success in the form of tensor transformations of $B^{\alpha}$
\cite{Abbott}.
In the case under consideration  we have
\beq
\label{B5}
\delta_{\omega}S_{gf}=
\int dx \sqrt{-{\bar g}}\;\big[\big(\delta_{\omega}B^{\alpha}+\omega^{\sigma}
\pa_{\sigma}B^{\alpha}\big)\chi_{\alpha}({\bar g},h)+
B^{\alpha}\omega^{\sigma}\pa_{\sigma}\chi_{\alpha}({\bar g},h)+B^{\alpha}
\delta_{\omega}\chi_{\alpha}({\bar g},h)\big].
\eeq

Suppose that the variation of gauge fixing functions $\chi_{\alpha}$
 under gauge transformations (\ref{B1}), (\ref{B3}) has
 the form
\beq
\label{B6}
\delta_{\omega}\chi_{\alpha}=
-\omega^{\sigma}\pa_{\sigma}\chi_{\alpha}-\pa_{\alpha}\omega^{\sigma}\chi_{\sigma},
\eeq
which corresponds to the transformation of vector fields of type $(0,1)$ (\ref{A5}).
Then choosing  the transformation law for $B^{\alpha}$ in the form
\beq
\label{B7}
\delta_{\omega}B^{\alpha}=-\omega^{\sigma}\pa_{\sigma}B^{\alpha}
+B^{\sigma}\pa_{\sigma}\omega^{\alpha},
\eeq
we arrive at the desired relation
\beq
\label{B8}
\delta_{\omega}S_{gf}=0.
\eeq
Notice that the transformation (\ref{B7}) coincides with the corresponding rule for tensor fields of type $(1,0)$, (\ref{A6}).

Due to the non-locality representation of the ghost action the its variation 
should be presented in detail
\beq
\nonumber
&&\delta_{\omega}S_{gh}=\int dxdydz\sqrt{-{\bar g}(x)}
\Big[\big( \delta_{\omega}{\bar C}^{\alpha}(x)+
\omega^{\sigma}(x)\pa_{\sigma}{\bar C}^{\alpha}(x)\big)G_{\alpha}^{\beta\gamma}(x,y)
R_{\beta\gamma\rho}(y,z)C^{\rho}(z)+\\
\nonumber
&&\qquad\qquad\qquad\qquad\qquad\qquad+{\bar C}^{\alpha}(x)\omega^{\rho}(x)\pa_{\rho}^x
G_{\alpha}^{\beta\gamma}(x,y)
R_{\beta\gamma\rho}(y,z)C^{\rho}(z)+
\\
\nonumber
&&\qquad\qquad\qquad\qquad\qquad\qquad+
{\bar C}^{\alpha}(x)G_{\alpha}^{\beta\gamma}(x,y)
R_{\beta\gamma\sigma}(y,z)\delta_{\omega}C^{\sigma}(z)+\\
\nonumber
&&\qquad\qquad\qquad\qquad\qquad\qquad+{\bar C}^{\alpha}(x)\delta_{\omega}G_{\alpha}^{\beta\gamma}(x,y)
R_{\beta\gamma\sigma}(y,z)C^{\sigma}(z)+\\
\label{B10}
&&\qquad\qquad\qquad\qquad\qquad\qquad+
{\bar C}^{\alpha}(x)G_{\alpha}^{\beta\gamma}(x,y)
\delta_{\omega}
\big(R_{\beta\gamma\sigma}(y,z)\big)C^{\sigma}(z)\Big].
\eeq
For variation of the gauge generators we find
\beq
\nonumber
&&\!\!\!\!\!\!\!\!\!\!\delta_{\omega}R_{\beta\gamma\sigma}(y,z)=\int\!\! dudv\Big[
\frac{\delta R_{\beta\gamma\rho}(y,v)}{\delta g_{\mu\nu}(u)}\omega^{\rho}(v)R_{\mu\nu\sigma}(u,z)-R_{\beta\gamma\lambda}(y,u)F^{\lambda}_{\sigma\rho}(u,z,v)\omega^{\rho}(v)\Big]=\\
\nonumber
&&\qquad\qquad\quad=-\omega^{\rho}(y)\pa_{\rho}^yR_{\beta\gamma\sigma}(y,z)-\pa_{\beta}\omega^{\rho}(y)R_{\gamma\rho\sigma}(y,z)-
\pa_{\gamma}\omega^{\rho}(y)R_{\beta\rho\sigma}(y,z)-\\
\label{B10b}
&&\qquad\qquad\qquad\qquad\qquad-R_{\beta\gamma\lambda}(y,z)\pa_{\sigma}\omega^{\lambda}(z)-
\pa_{\rho}^z\big(R_{\beta\gamma\sigma}(y,z)\omega^{\rho}(z)\big).
\eeq
The gauge transformation of $R_{\beta\gamma\sigma}(y,z)$ by itself differs
of the transformation law for the rensor field of type $(0,3)$.  It is no wonder
because of its  non-locality nature but it differs as well of the tensor transformatios of product of two tensors like $A_{\beta\gamma}(x)B_{\sigma}(y)$.

Having in mind the conditions (\ref{B6}) we can study a variation of the operator
(\ref{A18}) under the gauge transformations (\ref{B1}) and (\ref{B3}).
The result looks like  more complicated than (\ref{B10b}) and takes the form
\beq
\nonumber
\delta_{\omega}G_{\alpha}^{\beta\gamma}(x,y)=
\frac{\delta\delta_{\omega}\chi_{\alpha}(x)}{\delta h_{\beta\gamma}(y)}-
\int dz\; G_{\alpha}^{\mu\nu}(x,z)\frac{\delta\delta_{\omega}h_{\mu\nu}(z)}{\delta h_{\beta\gamma}(y)}-
\int dz\;\frac{\delta G_{\alpha}^{\beta\gamma}(x,y)}{\delta h_{\mu\nu}(z)}
\delta_{\omega}h_{\mu\nu}(z),
\eeq
or
\beq
\nonumber
&&\delta_{\omega}G_{\alpha}^{\beta\gamma}(x,y)=
-\omega^{\sigma}(x)\pa_{\sigma}^xG_{\alpha}^{\beta\gamma}(x,y)-\pa_{\alpha}\omega^{\sigma}(x)G_{\sigma}^{\beta\gamma}(x,y)+\\
\nonumber
&&\qquad+
G^{\beta\sigma}_{\alpha}(x,y)\pa_{\sigma}\omega^{\gamma}(y)
+G_{\alpha}^{\sigma\gamma}(x,y) \pa_{\sigma}\omega^{\beta}(y)-
\pa_{\sigma}^y\big(G_{\alpha}^{\beta\gamma}(x,y)\omega^{\sigma}(y)\big)-\\
\label{B9}
&&\qquad\qquad\qquad\qquad\qquad
-\int dz\;\frac{\delta G_{\alpha}^{\beta\gamma}(x,y)}{\delta h_{\mu\nu}(z)}
\delta_{\omega}h_{\mu\nu}(z).
\eeq
This transformations are again a far from the tensor transformation of type $(1,2)$ and of the product of tensors like $A_{\alpha}(x)B^{\beta\gamma}(y)$.

In the case of linear gauge fixing functions $\chi_{\alpha}$,
\beq
\label{B10}
\frac{\delta G_{\alpha}^{\beta\gamma}(x,y)}{\delta h_{\mu\nu}(z)}=0,
\eeq
the transformations (\ref{B9}) is simplified and 
we have

\beq
\nonumber
&&\delta_{\omega}S_{gh}=\int dxdydz\sqrt{-{\bar g}(x)}
\Big[\big( \delta_{\omega}{\bar C}^{\alpha}(x)+
\omega^{\sigma}(x)\pa_{\sigma}{\bar C}^{\alpha}(x)\big)G_{\alpha}^{\beta\gamma}(x,y)
R_{\beta\gamma\rho}(y,z)C^{\rho}(z)+\\
\nonumber
&&\qquad\qquad\qquad\qquad\qquad\qquad+{\bar C}^{\alpha}(x)\omega^{\rho}(x)\pa_{\rho}^x
G_{\alpha}^{\beta\gamma}(x,y)
R_{\beta\gamma\rho}(y,z)C^{\rho}(z)+
\\
\nonumber
&&\qquad\qquad\qquad\qquad\qquad\qquad+
{\bar C}^{\alpha}(x)G_{\alpha}^{\beta\gamma}(x,y)
R_{\beta\gamma\sigma}(y,z)\delta_{\omega}C^{\sigma}(z)-\\
\nonumber
&&\qquad\qquad\qquad\qquad\qquad\qquad
-{\bar C}^{\alpha}(x)\omega^{\rho}(x)\pa_{\rho}^x
G_{\alpha}^{\beta\gamma}(x,y)
R_{\beta\gamma\rho}(y,z)C^{\rho}(z)-\\
\nonumber
&&\qquad\qquad\qquad\qquad\qquad\qquad-
{\bar C}^{\alpha}(x)\pa_{\alpha}\omega^{\rho}(x)
G_{\rho}^{\beta\rho}(x,y)
R_{\beta\gamma\sigma}(y,z)C^{\sigma}(z)+\\
\nonumber
&&\qquad\qquad\qquad\qquad\qquad\qquad
+{\bar C}^{\alpha}(x)G_{\alpha}^{\beta\rho}(x,y)
\pa_{\rho}\omega^{\gamma}(y)R_{\beta\gamma\sigma}(y,z)C^{\sigma}(z)+\\
\nonumber
&&\qquad\qquad\qquad\qquad\qquad\qquad
+{\bar C}^{\alpha}(x)G_{\alpha}^{\rho\gamma}(x,y)
\pa_{\rho}\omega^{\beta}(y)R_{\beta\gamma\sigma}(y,z)C^{\sigma}(z)+\\
\nonumber
&&\qquad\qquad\qquad\qquad\qquad\qquad
+{\bar C}^{\alpha}(x)G_{\alpha}^{\beta\gamma}(x,y)
\omega^{\rho}(y)\pa_{\rho}^yR_{\beta\gamma\sigma}(y,z)C^{\sigma}(z)-\\
\nonumber
&&\qquad\qquad\qquad\qquad\qquad\qquad
-{\bar C}^{\alpha}(x)G_{\alpha}^{\beta\gamma}(x,y)
\omega^{\rho}(y)\pa_{\rho}^yR_{\beta\gamma\sigma}(y,z)C^{\sigma}(z)-\\
\nonumber
&&\qquad\qquad\qquad\qquad\qquad\qquad
-{\bar C}^{\alpha}(x)G_{\alpha}^{\beta\gamma}(x,y)
\pa_{\beta}\omega^{\rho}(y)R_{\gamma\rho\sigma}(y,z)C^{\sigma}(z)-\\
\nonumber
&&\qquad\qquad\qquad\qquad\qquad\qquad
-{\bar C}^{\alpha}(x)G_{\alpha}^{\beta\gamma}(x,y)
\pa_{\gamma}\omega^{\rho}(y)R_{\beta\rho\sigma}(y,z)C^{\sigma}(z)-\\
\nonumber
&&\qquad\qquad\qquad\qquad\qquad\qquad
-{\bar C}^{\alpha}(x)G_{\alpha}^{\beta\gamma}(x,y)
R_{\beta\gamma\lambda}(y,z)\pa_{\sigma}\omega^{\lambda}(z)C^{\sigma}(z)+\\
&&\qquad\qquad\qquad\qquad\qquad\qquad+{\bar C}^{\alpha}
(x)G_{\alpha}^{\beta\gamma}(x,y)
R_{\beta\gamma\sigma}(y,z)\omega^{\rho}(z)\pa_{\rho}C^{\sigma}(z)\Big].
\eeq
Finally
\beq
\nonumber
&&\delta_{\omega}S_{gh}=\int dxdydz\sqrt{-{\bar g}(x)}
\Big[\big( \delta_{\omega}{\bar C}^{\alpha}(x)+
\omega^{\sigma}(x)\pa_{\sigma}{\bar C}^{\alpha}(x)-\\
\nonumber
&&\qquad\qquad\qquad\qquad\qquad\qquad
-{\bar C}^{\rho}\pa_{\rho}\omega^{\alpha}(x)\big)G_{\alpha}^{\beta\gamma}(x,y)
R_{\beta\gamma\rho}(y,z)C^{\rho}(z)+\\
&&\qquad
+{\bar C}^{\alpha}(x)G_{\alpha}^{\beta\gamma}(x,y)
R_{\beta\gamma\sigma}(y,z)\big(\delta_{\omega}C^{\sigma}(z)+
\omega^{\rho}(z)\pa_{\rho}C^{\sigma}(z)-\pa_{\rho}\omega^{\sigma}(z)C^{\rho}(z)\big)\Big].
\eeq
Choosing the tensor transformation law for the ghost fields ${\bar C}^{\alpha}, C^{\alpha}$
\beq
&&\delta_{\omega}{\bar C}^{\alpha}(x)=
-\omega^{\sigma}(x)\pa_{\sigma}{\bar C}^{\alpha}(x)+
{\bar C}^{\rho}\pa_{\rho}\omega^{\alpha}(x),\\
&&\delta_{\omega}C^{\alpha}(x)=
-\omega^{\sigma}(x)\pa_{\sigma}C^{\alpha}(x)+
C^{\rho}\pa_{\rho}\omega^{\alpha}(x),
\eeq
we arrive at the invariance of the ghost action
\beq
\delta_{\omega}S_{gh}=0.
\eeq

Finally we conclude that the Faddeev-Popov action $S_{FP}$,
\beq
\label{B26}
\delta_{\omega}S_{FP}=0,
\eeq
is invariant under the background transformations of all fields $\phi,{\bar g}$,
\beq
\label{B27}
&&\delta^{(c)}_{\omega}{\bar g}_{\mu\nu}=R_{\mu\nu\sigma}({\bar g})\omega^{\sigma},\quad
\delta_{\omega}h_{\mu\nu}=R_{\mu\nu\sigma}(h)\omega^{\sigma},\\
\label{B28}
&&
\delta_{\omega}B^{\alpha}=-\omega^{\sigma}\pa_{\sigma}B^{\alpha}
+B^{\sigma}\pa_{\sigma}\omega^{\alpha},\quad
\delta_{\omega}{\bar C}^{\alpha}=-\omega^{\sigma}\pa_{\sigma}{\bar C}^{\alpha}
 +{\bar C}^{\sigma}\pa_{\sigma}\omega^{\alpha},\\
 &&
\delta_{\omega}C^{\rho}=-\omega^{\sigma}\pa_{\sigma}C^{\rho}+
\pa_{\sigma}\omega^{\rho}C^{\sigma}.
\label{B29}
\eeq
As the consequence of (\ref{B26}) the gauge invariance of the  vacuum functional follows
\beq
\delta_{\omega}Z({\bar g})=0.
\eeq
The same statement is valid for background effective action $\Gamma({\bar g})=\Gamma (\Phi=0,{\bar g})$
\beq
\delta_{\omega}\Gamma ({\bar g})=0.
\eeq

We see that the gauge invariance
for quantum gravity theories in the background field formalism can be achieved
if the two essential propositions related to  the transformation
law for gauge fixing functions (\ref{B6}) and to the linearity of these functions.  
If the gauge fixing functions are not linear in quantum fields $h_{\mu\nu}$,
\beq
\frac{\delta^2 \chi_{\alpha}(x)}{\delta h_{\beta\gamma}(y)\delta h_{\mu\nu}(z)}
\neq 0,
\eeq
then the tensor transformations (\ref{B27})-(\ref{B29}) cannot cancel the additional contribution (\ref{B9}) appearing in the variation of the ghost action
$S_{{\rm gh}}$.
Fulfilment or not fulfilment
of these requirements is closely related to a choice of gauge fixing functions
$\chi_{\alpha}=\chi_{\alpha}({\bar g},h)$. 

\section{Special choice of gauge fixing condition}
\noindent
A standard choice of $\chi_{\alpha}(\phi,{\bar g})$
corresponding to the background field
gauge condition  \cite{Barv} reads
\beq
\label{C1}
\chi_{\alpha}({\bar g}, h)=-{\bar g}^{\mu\lambda}\big(a
{\bar \nabla}_{\lambda}h_{\mu\alpha}+b{\bar \nabla}_{\alpha}h_{\mu\lambda}\big ),
\eeq
where ${\bar \nabla}_{\sigma}$ is the covariant derivative corresponding
the external metric tensor
${\bar g}_{\mu\nu}$ and $a,b$ are constants. 
The popular de Donder gauge condition
corresponds to the case when $a=1,b=-1/2$.

The choice (\ref{C1}) corresponds to linear dependence on quantum fields $h_{\mu\nu}$ so that we need to check 
the transformation law (\ref{B6}) only. 
The $\chi_{\alpha}=\chi_{\alpha}({\bar g}, h)$
are point functions of space-time coordinates $x$, $\chi_{\alpha}=\chi_{\alpha}(x)$,
 constructed with the help of second-rank
tensor fields ${\bar g}^{\mu\lambda}$ of type $(2,0)$ and third-rank tensor fields
${\bar \nabla}_{\lambda}h_{\mu\alpha}$ of type $(0,3)$ by contracting indices $\mu,\lambda$.
Therefore $\chi_{\alpha}(x)$ (\ref{C1}) is the tensor field of type $(0,1)$
with transformation law (\ref{A5}) that confirms the  transformation proposed (\ref{B6}).
The same result can be obtained by explicit calculations of gauge variation
of functions (\ref{C1}). We demonstrate this fact in the simplest case of a choice of
$\chi_{\alpha}$ when $a=0, b=-1$ so that
\beq
\label{C2}
\chi_{\alpha}({\bar g}, h)={\bar g}^{\mu\lambda}{\bar \nabla}_{\alpha}h_{\mu\lambda}=
{\bar \nabla}_{\alpha}\big({\bar g}^{\mu\lambda}h_{\mu\lambda}\big)=
\pa_{\alpha}\big({\bar g}^{\mu\lambda}h_{\mu\lambda}\big).
\eeq
Consider the gauge variation of (\ref{C2})
\beq
\label{C3}
\delta_{\omega}\chi_{\alpha}({\bar g}, h)=
\pa_{\alpha}\big(\delta^{(c)}_{\omega}{\bar g}^{\mu\lambda}\big)h_{\mu\lambda}+
\delta^{(c)}_{\omega}{\bar g}^{\mu\lambda}\pa_{\alpha}h_{\mu\lambda}+
+\pa_{\alpha}{\bar g}^{\mu\lambda}\delta_{\omega}h_{\mu\lambda}+
{\bar g}^{\mu\lambda}\pa_{\alpha}\big(\delta_{\omega}h_{\mu\lambda}\big),
\eeq
where
\beq
\label{C4}
\delta^{(c)}_{\omega}{\bar g}^{\mu\lambda}=
-{\bar g}^{\mu\alpha}\big(\delta^{(c)}_{\omega}{\bar g}_{\alpha\beta}\big){\bar g}^{\beta\lambda}=
-\omega^{\sigma}\pa_{\sigma}{\bar g}^{\mu\lambda}+
{\bar g}^{\sigma\lambda}\pa_{\sigma}\omega^{\mu}+
{\bar g}^{\mu\sigma}\pa_{\sigma}\omega^{\lambda},
\eeq
and $\delta_{\omega}h_{\mu\lambda}$ is given in (\ref{B3}). The set of terms in (\ref{C3})
 without derivatives of functions $\omega^{\sigma}$ is
\beq
\nonumber
&&-\omega^{\sigma}\big(h_{\mu\lambda}\pa_{\alpha}\pa_{\sigma}{\bar g}^{\mu\lambda}+
\pa_{\sigma}{\bar g}^{\mu\lambda}\pa_{\alpha}h_{\mu\lambda}
+\pa_{\alpha}{\bar g}^{\mu\lambda}\pa_{\sigma}h_{\mu\lambda}+
{\bar g}^{\mu\lambda}\pa_{\alpha}\pa_{\sigma}h_{\mu\lambda}\big)=\\
\label{C5}
&&\qquad\qquad\qquad=-\omega^{\sigma}\pa_{\alpha}\pa_{\sigma}
\big({\bar g}^{\mu\lambda}h_{\mu\lambda}\big)=
-\omega^{\sigma}\pa_{\sigma}\chi_{\alpha}.
\eeq
As to terms containing the second derivatives of $\omega$ in (\ref{C3}) we have
\beq
\label{C6}
{\bar g}^{\sigma\lambda}h_{\mu\lambda}\pa_{\alpha}\pa_{\sigma}\omega^{\mu}+
{\bar g}^{\mu\sigma}h_{\mu\lambda}\pa_{\alpha}\pa_{\sigma}\omega^{\lambda}-
{\bar g}^{\mu\lambda}h_{\mu\sigma}\pa_{\alpha}\pa_{\lambda}\omega^{\sigma}-
{\bar g}^{\mu\lambda}h_{\sigma\lambda}\pa_{\alpha}\pa_{\mu}\omega^{\sigma}=0.
\eeq
Collection of terms of the structure $\pa{\bar g}\pa\omega h$ in (\ref{C3}) reads
\beq
\nonumber
&&-\pa_{\sigma}{\bar g}^{\mu\lambda}\pa_{\alpha}\omega^{\sigma}h_{\mu\lambda}+
\pa_{\alpha}{\bar g}^{\mu\sigma}\pa_{\sigma}\omega^{\lambda}h_{\mu\lambda}+
\pa_{\alpha}{\bar g}^{\sigma\lambda}\pa_{\sigma}\omega^{\mu}h_{\mu\lambda}-\\
\label{C7}
&&-\pa_{\alpha}{\bar g}^{\mu\lambda}\pa_{\lambda}\omega^{\sigma}h_{\mu\sigma}-
\pa_{\sigma}{\bar g}^{\mu\lambda}\pa_{\mu}\omega^{\sigma}h_{\sigma\lambda}=-
\pa_{\sigma}{\bar g}^{\mu\lambda}\pa_{\alpha}\omega^{\sigma}h_{\mu\lambda}.
\eeq
In its turn the terms of the structure ${\bar g}\pa\omega \pa h$  enter in (\ref{C3})
in the form
\beq
\nonumber
&&{\bar g}^{\mu\sigma}\pa_{\sigma}\omega^{\lambda}\pa{\alpha}h_{\mu\lambda}-
{\bar g}^{\mu\lambda}\pa_{\alpha}\omega^{\sigma}\pa_{\sigma}h_{\mu\lambda}-
{\bar g}^{\mu\lambda}\pa_{\lambda}\omega^{\sigma}\pa_{\alpha}h_{\mu\sigma}-\\
\label{C8}
&&-{\bar g}^{\mu\lambda}\pa_{\lambda}\pa_{\mu}\omega^{\sigma}\pa_{\alpha}h_{\sigma\lambda}+
{\bar g}^{\sigma\lambda}\pa_{\sigma}\omega^{\mu}\pa_{\alpha}h_{\mu\sigma}=-
{\bar g}^{\sigma\lambda}\pa_{\alpha}\omega^{\sigma}\pa_{\sigma}h_{\mu\lambda}.
\eeq
Finally we have the result
\beq
\nonumber
&&\delta_{\omega}\chi_{\alpha}=
-\omega^{\sigma}\pa_{\sigma}\chi_{\alpha}
-\pa_{\sigma}{\bar g}^{\mu\lambda}\pa_{\alpha}\omega^{\sigma}h_{\mu\lambda}
-{\bar g}^{\sigma\lambda}\pa_{\alpha}\omega^{\sigma}\pa_{\sigma}h_{\mu\lambda}=\\
\label{C9}
&&\qquad\;\;=-\omega^{\sigma}\pa_{\sigma}\chi_{\alpha}-
\pa_{\alpha}\omega^{\sigma}\pa_{\sigma}\big({\bar g}^{\mu\lambda}h_{\mu\lambda}\big)=
-\omega^{\sigma}\pa_{\sigma}\chi_{\alpha}-\pa_{\alpha}\omega^{\sigma}\chi_{\sigma},
\eeq
which confirms the transformations (\ref{B6}). In a similar way one can check the rightness
of (\ref{B6}) for (\ref{C1}) when $a\neq 0$ but corresponding calculations look more complicated
due to the covariant derivative ${\bar \nabla}_{\lambda}$ and we omit them.

\section{Discussion}
\noindent
In the present paper we have considered the background field formalism for
Quantum Gravity from point of view of choice of the gauge fixing condition.  
Application of this formalism to the Yang-Mills theories is very effective means in quantum region
(among recent investigations see, for example, \cite{Barv,BLT-YM})
because it allows to support
 gauge invariance on all stages of calculations. The quantum gravity theories look like as special type of gauge theories of Yang-Mills fields
with closed gauge algebra and with structure coefficients independent on fields
and therefore they can be quantized in the form of the Faddeev-Popov procedure. Then for all admissable choice of gauge condition both the vacuum functional and the background effective action on its extremals are gauge independent. The property of gauge invariance of the vacuum functional and the backfground effective action is more sensitive to the choice of gauges.
It has been verified explicitly that the gauge invariance can be arrived at the fulfilment of two conditions: a) the linearity of gauge fixing functions with respect to quantum gravitational fields  and b) the tensor transformations for 
gauge fixing functions under the background gauge transformations.

\section*{Acknowledgments}
\noindent
 The author thanks I.L. Buchbinder and I.V. Tyutin for useful discussions. The work is supported in part by the
Ministry of Education and Science of the Russian Federation, grant
3.1386.2017 and by the RFBR grant 18-02-00153.

\end{document}